\documentclass[a4paper]{jpconf}
\usepackage{graphicx}
\usepackage[latin1]{inputenc}
\usepackage{amsmath}
\usepackage{amsfonts}
\usepackage{rotating}
\usepackage{amssymb}

\newtheorem{teorema}{Teorema}[section]

\newtheorem{theorem}[teorema]{Theorem}

\begin{document}
\title{Complete classification of second-order symmetric spacetimes}

\author{O F Blanco$^{1}$, M S\'anchez$^{1}$ and J M M Senovilla$^2$}

\address{$^{1}$ Departamento de Geometr\'ia y Topolog\'ia , Facultad de Ciencias, Universidad de Granada
Campus Fuentenueva s/n, 18071 Granada, Spain
}
\address{$^2$ F\'isica Te\'orica, Universidad del Pa\'is Vasco, Apartado 644, 48080 Bilbao, Spain}

\ead{oihane@ugr.es, sanchezm@ugr.es, josemm.senovilla@ehu.es}

\begin{abstract}
We present the explicit local form of the metric of the second-order symmetric non-symmetric 4-dimensional Lorentzian manifolds. They turn out to be a
specific subclass of plane waves.

\end{abstract}

\section{Introduction}

The historical roots of the subject under investigation go back to the classification of the locally symmetric spaces in the 
(proper) Riemannian case by Cartan \cite{CA,CAbis}. {\it Locally symmetric} pseudo-Riemannian manifolds 
 are characterized by the condition
 $\nabla R=0$, where $R$ denotes the curvature tensor of the manifold. 
The pseudo-Riemannian manifolds satisfying $\nabla^{2}R=0$ constitute
 a logical generalization of the locally symmetric ones. But in the Riemannian
 case these two conditions are equivalent. 
Moreover, in the Riemannian case:

\begin{equation}\label{1}
\nabla^{m}R=0, m\geq 2\Longrightarrow \nabla R=0, 
\end{equation}

\noindent a result based on the orthogonal de Rham decomposition \cite{NO,TA}. In this Riemannian case
 the natural generalization are the semi-symmetric 
spaces (introduced in \cite{CA2}, studied in \cite{SZ1,SZ2}) defined by the relation 
${\cal R}_{XY}R=0$ for all vector fields $X,Y$, where ${\cal R}_{XY}$ is the curvature operator $\nabla_{X}\nabla_{Y}-\nabla_{Y}\nabla_{X}-\nabla_{[X,Y]}$.



In the Lorentzian case (\ref{1}) does not hold, because of the failure of the 
orthogonal de Rham decomposition 
\cite{WU1, WU2, WU3}. In fact, 
\begin{center}
     $\nabla^{m}R=0, m\geq 2\Longrightarrow g(\nabla^{m-1} R,\nabla^{m-1} R)=0$,
        \end{center}

\noindent where the metric $g$ is extended to the inner product between tensor fields in the standard fashion. So in
 the Lorentzian case we can introduce the {\it second-order symmetric} (also called {\it 2-symmetric}) spaces \cite{SN1}, i.e., the spaces that 
satisfy the condition $\nabla^{2}R=0$. 

The following relations hold: $\{\nabla R=0\}\subset\{\nabla^{2}R=0\}\subset\{{\cal R}_{XY} R=0\}.$

In this contribution we present a specific result from \cite{SN2}. Namely, we find the 
2-symmetric non-symmetric 4-dimensional spacetimes: the explicit local form of 
the metric, classification and properties. The main result is summarized in the following theorem:

\begin{theorem}\label{t3}
A 2-symmetric non-symmetric 4-dimensional spacetime $(M,g)$ is locally isometric to $\mathbb{R}^4$ endowed with the metric 
$$ds^2=-2 du (dv+H du)+dx^2+dy^2,$$ 
\noindent where $H(u,x,y)=(\alpha_1 u+\beta_1)x^2+(\alpha_2 u +\beta_2)y^2 +(\alpha_3 u + \beta_3)x y$ for some constants 
$\{\alpha_A,\beta_A\}_{A=1,2,3}$ with $\alpha_1^2+\alpha_2^2+\alpha_3^2\neq 0$.
\vspace{0.2cm}

\end{theorem}

\section{Sketch of the proof of theorem \ref{t3}}

The starting point for this proof is theorem 4.2 in \cite{SN1}. From this theorem follows that a 2-symmetric non-symmetric spacetime must have a null parallel vector field and 
therefore its metric is of Brinkmann type:
$$
ds^2=-2 du (dv+H du+W_i dx^i)+g_{ij}dx^idx^j,\hspace{0.2cm} i,j\in\{2,3\},
$$
\noindent where $H$, $W_i$ and $g_{ij}=g_{ji}$ are functions independent of $v$, otherwise arbitrary.

Using the 2-symmetry condition, one can prove that $g_{ij}$ must be locally symmetric on each slice $u=u_0$ \cite{SN2}.
 Since the only locally 
symmetric 2-dimensional Riemannian spaces are the constant curvature ones, we have that in appropriate local coordinates
$$g_{ij}dx^idx^j=\frac{1}{\left( 1+\frac{k(u)}{4}\left( x^2+y^2\right)\right) ^2}\left( dx^2+dy^2\right).$$

In fact, $k(u)$ must vanish. To prove this, we use the Petrov classification \cite{ES}. By \cite{IE} we know that all the semi-symmetric spacetimes (of dimension $4$) are of type
$D$, $N$ or $O$. Moreover, as a consequence of this result,
 it was also proven that the 2-symmetric spacetimes are of Petrov type $N$, or its degenerate case,
 $O$. And this only happens when $k(u)=0$.

Hence, the line element of a 2-symmetric non-symmetric 4-dimensional spacetime $(M,g)$ simplifies to 
$$ds^2=-2 du (dv+H du+W_2 dx+W_3 dy)+dx^2+dy^2,  
$$

From now on, an overdot means derivative with respect to $u$, $f_{,x}=\partial f/\partial x$
 and $f_{,y}=\partial f/\partial y$. 

Using the Cartan method (see, for example, \cite{CH}) to calculate the connection 1-forms and
 the curvature 2-forms for the null coframe $
\{\theta^{0}, \theta^{1}, \theta^{2}, \theta^{3}\}=\{du, dv + H du+W_2 dx+W_3 dy, dx, dy\}$
and solving the 2-symmetry equations in this 
frame, we conclude that there exist functions $\{a_{rs}(u),b_{rs}(u)\}_{\{r,s\}\in\{0,1\},r+s<2}$, $\omega(u,x,y)$, $w_2(u)$, 
$w_3(u)$ and 
$r(u)$ such that \begin{equation}\label{ww2}W_2(u,x,y)=\omega_{,x}(u,x,y)+w_2(u) y,
\end{equation}
\begin{equation}\label{ww3}W_3(u,x,y)=\omega_{,y}(u,x,y)+w_3(u) x,
\end{equation}
\begin{equation}\label{H3}
H(u,x,y)=\frac{a_{10}}{2}x^2+\frac{b_{01}}{2}y^2+b_{10}xy +a_{00}x+b_{00}y+r(u)+Q(u,x,y)
\end{equation}
\noindent where $Q_{,y}(u,x,y)=\dot{W}_3(u,x,y)$, and the following equations must be satisfied:
\begin{equation}\label{2s1}
 -\ddot{a}_{10}+ 2 g \ddot{g}+2 \dot{g}a_{01}-2g^2(b_{01}-a_{10})+4 g \dot{a}_{01}=0,
\end{equation}
\begin{equation}\label{2s2}
 \ddot{b}_{01}-2 g \ddot{g}+2\dot{g}b_{10}+4g\dot{b}_{10}-2g^2(b_{01}-a_{10})=0,
\end{equation}
\begin{equation}\label{2s3}
  \ddot{a}_{01}+ \dddot{g}-4\dot{g}g^2+\dot{g}(a_{10}-b_{01})-2g^2a_{01}-2 g (\dot{a}_{10}-\dot{b}_{01})=0,
\end{equation}
\begin{equation}\label{2s4}
  b_{10}-a_{01}=2\dot{g},
\end{equation}
\begin{equation}\label{2s5}
 g(u)=\frac{1}{2}(W_{2,y}-W_{3,x}).
\end{equation}


The proof can then be completed by noticing that Eqs.(\ref{ww2}), (\ref{ww3}) and 
(\ref{H3}) together with (\ref{2s1})--(\ref{2s5}) permit the coordinate change needed to arrive at the final form of the metric. This change of coordinates is given by
$$u'=u+u_0,
$$

$$v'=v+F(u,x,y),
$$

$$x'=\cos\left(\theta(u)\right)x-\sin\left(\theta(u)\right)y+B_2(u),
$$

$$y'=\sin(\theta(u))x+\cos\left(\theta(u)\right)y+B_3(u),
$$

It is easily verified that the functions $\theta$, $B_2$, $B_3$ and $F$ can be chosen such that the line-element becomes
\begin{equation}\label{me3}
ds'^2=-2 du'(dv'+H'du')+(dx')^2+(dy')^2,
\end{equation}
\noindent where 
\begin{equation} \label{H5}
H'(u',x',y')=p_1(u') (x')^2+ p_2(u') (y')^2+p_3(u') x'y'
\end{equation}
\noindent for some new functions $\{p_A(u')\}_{A=1,2,3}$, because the integrability conditions of the resulting differential equations are always met by virtue of equations (\ref{ww2})--(\ref{2s5}).

To finalize the proof of theorem \ref{t3}, one simply has to solve the equations (\ref{2s1})-(\ref{2s5}) appropriately restricted to the new form of the metric (\ref{me3}) with (\ref{H5}). This readily implies that $p_1(u')$, $p_2(u')$ and $p_3(u')$ must be linear functions: 
$$p_A(u')=\alpha_A u'+\beta_A,~\alpha_A, \beta_A\in\mathbb{R}, ~A\in\{1,2,3\}.
$$


The condition $\alpha_1^2+\alpha_2^2+\alpha_3^2\neq 0$ must be enforced as otherwise $\nabla R=0$ and the spacetime would be locally symmetric.

\section {Concluding remarks}
We would like to stress that, even though the 
family of metrics obtained in Therorem \ref{t3} depend on six parameters, a further analysis shows that only four of them are essential parameters, and the other two can be removed by the remaining freedom of coordinates.

Similarly, we want to remark that our result is local. A global classification is also feasible, though some global hypotheses on the spacetime seem unavoidable to that end. Such kind of
hypotheses may involve the topology of the manifold, geodesics or causality (see \cite{CFS,FS,FS2} for a  study of the last two issues in some wave-type spacetimes). In particular, it seems that the local solutions obtained in Theorem \ref{t3} are also the unique global 4-dimensional solutions under the requirements of simple connectedness and geodesic completeness.

An open question is whether or not the above pattern is maintained for higher-order symmetric 4-dimensional spacetimes. Specifically, we wonder if the $m$-symmetric spacetimes (those with $\nabla^m R=0$) are given, for general $m$, precisely by the line-element  (\ref{me3}) with (\ref{H5}) where the functions $p_1(u')$, $p_2(u')$ and $p_3(u')$ are polynomials of degree $m$.

A more important open question is the extension of the result to higher dimensions. Of course, many more possibilities appear in this general case, which are 
being analyzed in \cite{SN2}. 

\ack
JMMS is supported by grants FIS2004-01626 (MICINN) and GIU06/37 (UPV/EHU). OFB and MS are partially 
supported by grants P06-FQM-01951 (J. Andaluc\'ia) and 
 MTM2007-60731 (Spanish
MEC with FEDER funds).

\end{document}